\documentclass[twocolumn,aps,prc,superscriptaddress,showpacs,floatfix,longbibliography]{revtex4-1}
\usepackage{url}
\usepackage{cancel}
\usepackage[colorlinks,linkcolor=blue,citecolor=blue,filecolor=black,urlcolor=blue]{hyperref}
\usepackage{epsfig,graphics}
\usepackage{graphicx}% Include figure files
\usepackage{dcolumn}% Align table columns on decimal point
\usepackage{bm}% bold math
\usepackage[usenames]{color}
\usepackage{amssymb}
\usepackage{amsmath}
\usepackage{multirow}
\usepackage{float}
\usepackage{harpoon}
\usepackage{MnSymbol}
\usepackage{appendix}
\usepackage{color}
\usepackage{hyperref}
\usepackage{cleveref}

\newcommand{\sqrtsnn}{\mbox{$\sqrt{s_{\mathrm{NN}}}$}}
\newcommand{\pT} {p_{\mathrm{T}}}

\newcommand{\nch}{N_{\mathrm{ch}}}

\begin{document}

\title{Probing neutron-skin thickness with free spectator neutrons in ultracentral high-energy isobaric collisions}
\author{Lu-Meng Liu}
\affiliation{School of Physical Sciences, University of Chinese Academy of Sciences, Beijing 100049, China}
\author{Chun-Jian Zhang}
\affiliation{Department of Chemistry, Stony Brook University, Stony Brook, NY 11794, USA}
\author{Jia Zhou}
\affiliation{Shanghai Institute of Applied Physics, Chinese Academy of Sciences, Shanghai 201800, China}
\affiliation{School of Physical Sciences, University of Chinese Academy of Sciences, Beijing 100049, China}
\author{Jun Xu}\email[Correspond to\ ]{xujun@zjlab.org.cn}
\affiliation{Shanghai Advanced Research Institute, Chinese Academy of Sciences, Shanghai 201210, China}
\affiliation{Shanghai Institute of Applied Physics, Chinese Academy of Sciences, Shanghai 201800, China}
\author{Jiangyong Jia}\email[Correspond to\ ]{jiangyong.jia@stonybrook.edu}
\affiliation{Department of Chemistry, Stony Brook University, Stony Brook, NY 11794, USA}
\affiliation{Physics Department, Brookhaven National Laboratory, Upton, NY 11976, USA}
\author{Guang-Xiong Peng}
\affiliation{School of Nuclear Science and Technology, University of Chinese Academy of Sciences, Beijing 100049, China}
\affiliation{Theoretical Physics Center for Science Facilities, Institute of High Energy Physics, Beijing 100049, China}
\affiliation{Synergetic Innovation Center for Quantum Effects $\&$ Applications, Hunan Normal University, Changsha 410081, China}
\date{\today}
\begin{abstract}
We show that the yield ratio of free spectator neutrons produced in high-energy $^{96}$Zr+$^{96}$Zr to $^{96}$Ru+$^{96}$Ru collisions is a clean probe of the neutron-skin thickness of colliding nuclei and the slope parameter $L$ of the symmetry energy. The idea is demonstrated based on the proton and neutron density distributions via a state-of-the-art Skyrme-Hartree-Fock-Bogolyubov calculation. Among spectator nucleons given by the Glauber model, free spectator neutrons include those from direct production that survive from clusterization as well as those from deexcitation of heavy clusters described by the popular GEMINI model. More free neutrons are produced in collisions of $^{96}$Zr nucleus due to its larger neutron skin, compared to those produced in collisions of $^{96}$Ru nucleus with a smaller neutron skin. The difference of the free spectator neutron yield is further increased with the increasing difference of the neutron-skin thickness between $^{96}$Zr and $^{96}$Ru with a larger $L$ value, and the increase in ultracentral collisions is particularly insensitive to model details and experimental uncertainties. Since the production of free spectator neutrons is not affected by the complicated dynamics in the mid-rapidity region, the ratio of their multiplicities in ultracentral isobaric collisions is a robust observable for constraining the neutron skin and $L$ value.
\end{abstract}
\maketitle

{\bf Introduction.} The distributions of protons and neutrons in heavy nuclei are the primary probes of the nuclear interaction as well as the nuclear matter equation of state (EOS)~\cite{Brown:2000pd,Typel:2001lcw}, particularly the symmetry energy that describes how the energy per nucleon changes with the neutron-proton asymmetry~\cite{Thiel:2019tkm}. In heavy neutron-rich nuclei, the excess neutrons are pushed out to form a neutron skin, with its thickness $\Delta r_{\mathrm{np}}$ defined as the difference between the neutron and proton root-mean-square (RMS) radii. The value of $\Delta r_{\mathrm{np}}$ is a robust probe of the slope parameter $L$ of the symmetry energy~\cite{Horowitz:2000xj,Furnstahl:2001un,Todd-Rutel:2005yzo,Centelles:2008vu,Zhang:2013wna,Xu:2020fdc}. Besides its relevance in nuclear structure, the measurement of $\Delta r_{\mathrm{np}}$ and the subsequent determination of $L$ are important in understanding many interesting phenomena in both astrophysics~\cite{Steiner:2004fi,Lattimer:2006xb} and nuclear reactions~\cite{Li:2008gp}.

In the past decades, various methods have been used to measure $\Delta r_{\mathrm{np}}$, including hadron-nucleus scatterings~\cite{Trzcinska:2001sy,Klos:2007is,Brown:2007zzc,Terashima:2008rb,Zenihiro:2010zz,Friedman:2012pa}, photon-nucleus scatterings~\cite{Tarbert:2013jze}, and, particularly, parity-violating electron-nucleus scatterings~\cite{Abrahamyan:2012gp,PREX:2021umo}, the data from which has induced considerable attention~\cite{Reed:2021nqk} and debate~\cite{Piekarewicz:2021jte,PhysRevC.105.055503}. Recently, it is realized that the huge amount of particles produced in high-energy heavy-ion collisions at $\sqrtsnn\gtrsim100$ GeV can be used to determine the density distribution of colliding nuclei~\cite{Filip:2009zz,Shou:2014eya,Giacalone:2019pca}, including their neutron skins~\cite{Li:2019kkh}. These collisions produce a hot dense matter known as the quark-gluon plasma (QGP), whose space-time evolution is well described by relativistic hydrodynamics. The latter in turn can be used to image the shape and radial structures of the colliding nuclei~\cite{Jia:2021tzt}. The best example for this possibility is illustrated by comparing $^{96}$Ru+$^{96}$Ru and $^{96}$Zr+$^{96}$Zr collisions at $\sqrtsnn=200$ GeV. The ratios of many observables between these two isobaric systems, published by the STAR Collaboration~\cite{STAR:2021mii}, show significant deviations from unity, each with its own characteristic centrality dependencies. Since isobar nuclei have the same mass number, these deviations must originate from the difference in the structure of the colliding nuclei, which impact the initial state of QGP and its final state observables. Model calculations show that the ratios of elliptic flow $v_2$ and triangular flow $v_3$ suggest a large quadrupole deformation $\beta_2$ in $^{96}$Ru and a large octupole deformation $\beta_3$ in $^{96}$Zr, respectively~\cite{Zhang:2021kxj}. In mid-central collisions, the ratio of multiplicity distribution and/or $v_2$ can also probe the $\Delta r_{\mathrm{np}}$ in $^{96}$Zr and $^{96}$Ru~\cite{Li:2019kkh} and/or their difference~\cite{Jia:2021oyt}. Predictions for many other observables and their sensitivities to the deformation and neutron skin have been made, such as the mean transverse momentum $\pT$~\cite{Xu:2021uar} and its fluctuations~\cite{PhysRevC.105.044905}, and $v_n$--$\pT$ correlations~\cite{Giacalone:2019pca,Bally:2021qys,Jia:2021wbq}.

\begin{figure}[!h]
\includegraphics[width=0.9\linewidth]{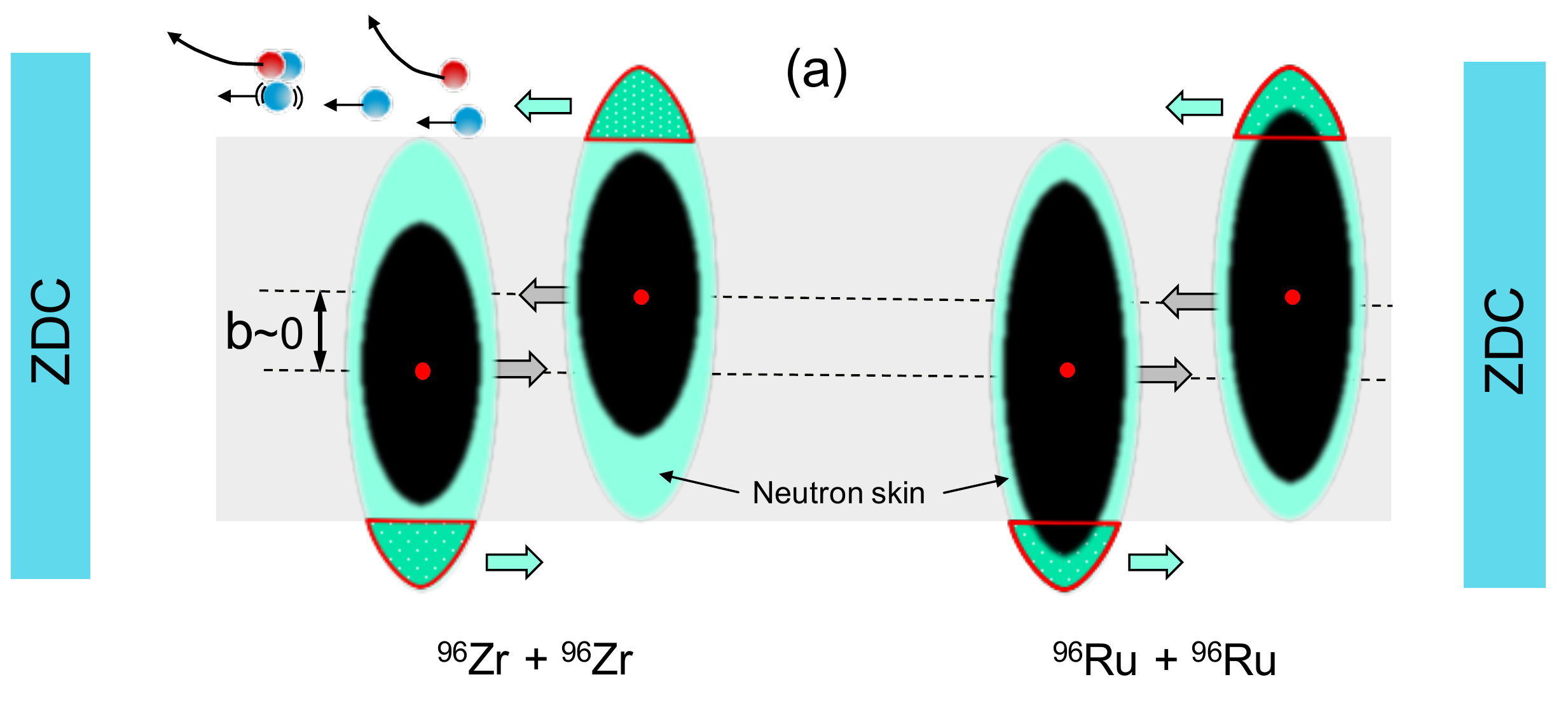}\\
\includegraphics[width=0.8\linewidth]{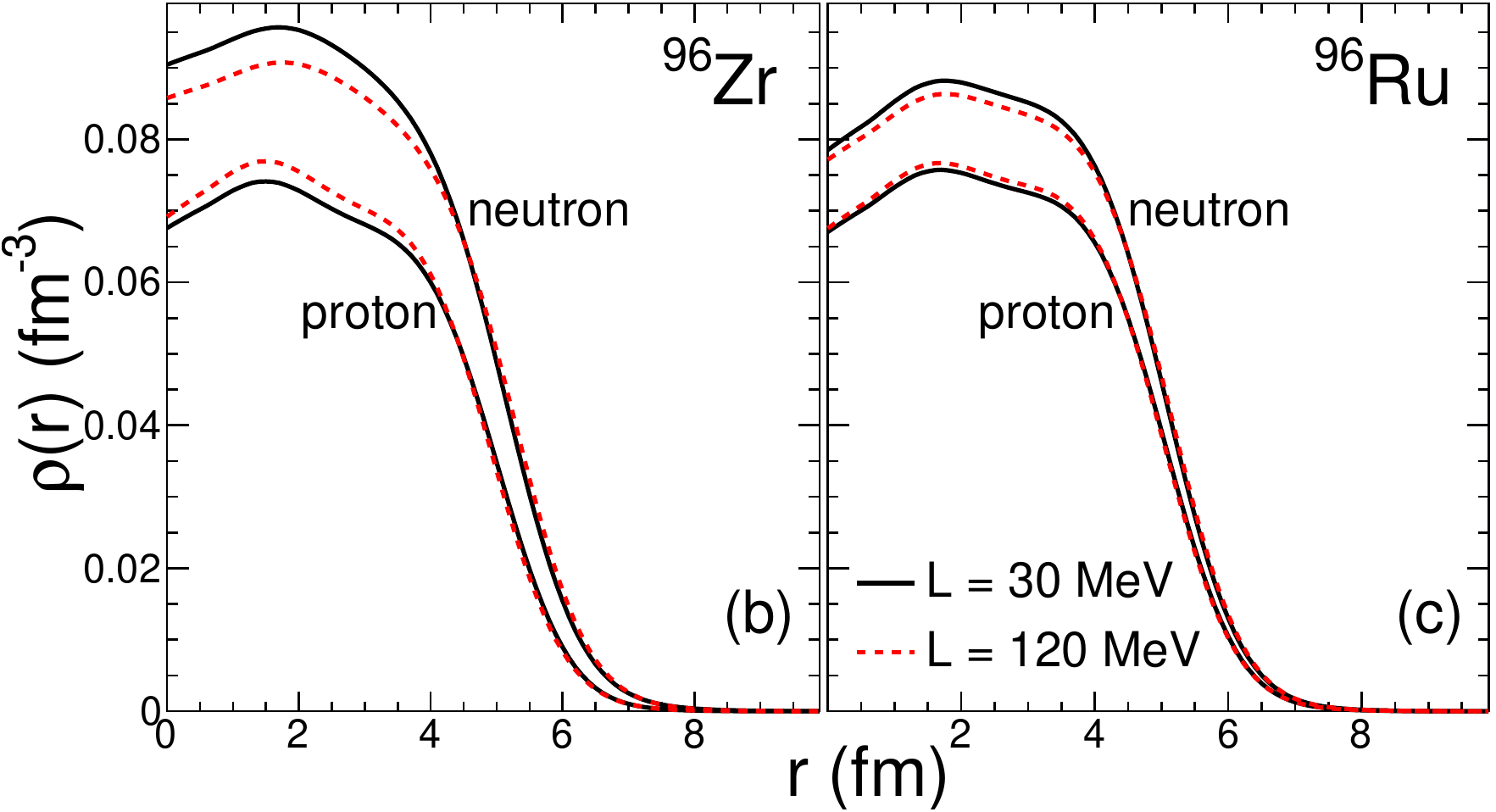}
\vspace*{-.3cm}
\caption{\label{fig1} (top) Side view of central collisions of $^{96}$Zr with a larger neutron skin and $^{96}$Ru with a smaller neutron skin. The directly-produced spectator neutrons and those from the deexcitation of charged clusters go to the ZDC down the beamline. The free protons and remaining charged clusters are bent away by beam optics. (bottom) Density profiles of neutrons and protons for $^{96}$Zr and $^{96}$Ru from spherical SHFB calculations using different slope parameters $L$ of the symmetry energy.}
\end{figure}

Most observables in heavy-ion collisions utilize particles produced near the mid-rapidity region, the description of which requires sophisticated modeling of the full space-time dynamics and properties of the QGP. In this Letter, we propose the ratio of the number of free spectator neutrons $N_n$ in isobar systems, easily measurable by the zero-degree calorimeters (ZDC)~\cite{STAR:2021mii}, as a clean and complementary probe of the neutron-skin thicknesses of $^{96}$Zr and $^{96}$Ru. In ultracentral collisions (UCC), most free spectator nucleons, outside the overlap region, originate from the surface of the nucleus (see cartoon in Fig.~\ref{fig1}). In high-energy heavy-ion collisions, these free neutrons, captured by the ZDC, are casually disconnected from the matter produced by the participating nucleons. The $N_n$ is sensitive to the $\Delta r_{\mathrm{np}}$, i.e., a larger $\Delta r_{\mathrm{np}}$ in $^{96}$Zr than in $^{96}$Ru is expected to give a larger $N_n$ in $^{96}$Zr+$^{96}$Zr collisions than in $^{96}$Ru+$^{96}$Ru collisions. The free spectator neutrons can be produced either directly or from the deexcitation of charged nucleon clusters. Despite being measured routinely in heavy-ion experiments, the distribution of $N_n$ is mostly used to estimate the event centrality or the reaction plane~\cite{PHENIX:2000owy,ALICE:2013hur}. Our study is the first to use the $N_n$ to probe the collective structure of colliding nuclei. The recent STAR paper~\cite{STAR:2021mii} has observed a significant enhancement of $N_n$ in $^{96}$Zr+$^{96}$Zr collisions relative to $^{96}$Ru+$^{96}$Ru collisions, providing a clear motivation for our study.

{\bf Method.} We generate the spatial distributions of neutrons and protons in the initial $^{96}$Zr and $^{96}$Ru using the Skyrme-Hartree-Fock (SHF) model~\cite{Chen:2010qx}, where the 10 parameters in the Skyrme interaction can be expressed analytically in terms of 10 macroscopic quantities including $L$. The model allows us to vary $L$ while keeping the other parameters fixed at their empirical values~\cite{Chen:2010qx} that give a reasonable description of global nuclear structure data. Based on the Skyrme interaction, the energy-density functional can then be obtained using the Hartree-Fock method, and the single-particle Hamiltonian is obtained using the variational principle. Solving the Schr\"odinger equation gives the wave functions of constituent neutrons and protons and thus their density distributions. As an essential ingredient for the description of open shell nuclei, the paring correlations have also been incorporated in above standard procedure, and this is called the Skyrme-Hartree-Fock-Bogolyubov (SHFB) method~\cite{Stoitsov:2012ri}. The effects of axial symmetry deformation are included in the SHFB calculation using the cylindrical transformed deformed harmonic oscillator basis. An analysis on the ratio of $v_2$ and $v_3$ in the isobaric collisions favors $\beta_2=0.16$ for $^{96}$Ru and $\beta_2=0.06$ and $\beta_3=0.20$ for $^{96}$Zr~\cite{Zhang:2021kxj}. We will compare results from spherical and deformed density distributions, and will show later that the results are relatively insensitive to the deformation parameters. The bottom panels of Fig.~\ref{fig1} show the density profiles of $^{96}$Zr and $^{96}$Ru from spherical SHFB calculations, from which we obtain $\Delta r_{\mathrm{np}}=0.147$ fm in $^{96}$Zr and 0.028 fm in $^{96}$Ru for $L=30$ MeV, and $\Delta r_{\mathrm{np}}=0.231$ fm in $^{96}$Zr and 0.061 fm in $^{96}$Ru for $L=120$ MeV, respectively. The influence of nuclear deformation on $\Delta r_{\mathrm{np}}$ for a given $L$ is found to be less than 8\%. The charge radii for $^{96}$Zr and $^{96}$Ru in different cases are consistent with the experimental data~\cite{Angeli:2004kvy} within $1.5\%$.

The $^{96}$Zr+$^{96}$Zr and $^{96}$Ru+$^{96}$Ru collisions are simulated using a Monte-Carlo Glauber model. In each collision event, the colliding nuclei are placed at a random impact parameter and their orientations are uniformly randomized. The coordinates of neutrons and protons are sampled according to the density distributions obtained from the SHFB calculation, while their momenta are sampled in the isospin-dependent Fermi sphere according to the local densities of neutrons and protons. The nucleon-nucleon inelastic cross section are chosen to be 42 mb at $\sqrt{s_{NN}}=200$ GeV. From this, the participant nucleons and spectator nucleons are identified. The quantities associated with participant matter, such as the number of participating nucleons  and nucleon-nucleon collisions, are calculated for each event. The distribution of these quantities are then used in a fit to the measured distribution of charged-particle multiplicity $\nch$ according to the experimental procedure~\cite{STAR:2021mii}. The quantities associated with spectator matter, in particular the $N_n$ to be determined below, can then be correlated with $\nch$, similar to that from the experimental analysis.

The spectator matter are further grouped into charged clusters and free nucleons. Nucleons close in phase space, i.e., with the distance $\Delta r<\Delta r_{\mathrm{max}}=3$~fm and relative momentum $\Delta p<\Delta p_{\mathrm{max}}=300$ MeV/$c$, are assigned to the same cluster~\cite{Li:1997rc}. Here $\Delta r_{\mathrm{max}}$ and $\Delta p_{\mathrm{max}}$ represent the scale of maximum distance and relative momentum between neighboring nucleons in the clusters, respectively.

The deexcitation of heavy clusters with $A \geq 4$ are handled by the GEMINI model~\cite{Charity:1988zz,Charity:2010wk}, which requires as inputs the angular momentum and the excitation energy of the cluster. The angular momentum of the cluster is calculated by summing those from all nucleons with respective to their center of mass (C.M.), while the energy of the cluster is calculated from a simplified SHF energy-density functional~\cite{Chen:2010qx}, with the neutron and proton phase-space information obtained from averaging over parallel events~\cite{Wong:1982zzb,Bertsch:1988ik} for the same impact parameter and collision orientation. Based on the same ground-state nucleon phase-space information, the binding energy per nucleon obtained from this method for most nuclei, including $^{96}$Ru and $^{96}$Zr, are found to agree with the SHFB calculation within $\pm1$ MeV, therefore justifying this approach. The excitation energy is then calculated by subtracting from the calculated cluster energy the ground-state energy of known nuclei taken from Ref.~\cite{Wang:2021xhn}. For clusters with compositions absent in Ref.~\cite{Wang:2021xhn}, an improved liquid-drop model~\cite{Wang:2014qqa} is employed to calculate their ground-state energies.

For spectator nucleons that do not form heavy clusters ($A \geq 4$), they could still coalesce into light clusters with $A \leq 3$, i.e., deuterons, tritons, and $^3$He, and the formation probabilities are calculated according to the following Wigner functions~\cite{Chen:2003ava,Sun:2017ooe} in their C.M. frame, i.e.,
\begin{eqnarray}
f_d &=& 8g_d \exp{\left(-\frac{\rho^2}{\sigma_d^2}-p_\rho^2\sigma_d^2\right)}, \\
f_{t/^3He} &=& 8^2 g_{t/^3\mathrm{He}} \exp{\left[-\left(\frac{\rho^2+\lambda^2}{\sigma_{t/^3\mathrm{He}}^2}\right)- (p_\rho^2+p_\lambda^2)\sigma_{t/^3\mathrm{He}}^2\right]},
\end{eqnarray}
with $\vec{\rho}=(\vec{r}_1-\vec{r}_2)/\sqrt{2}$, $\vec{p}_\rho=(\vec{p}_1-\vec{p}_2)/\sqrt{2}$, $\vec{\lambda}=(\vec{r}_1+\vec{r}_2-2\vec{r}_3)/\sqrt{6}$, and $\vec{p}_\lambda=(\vec{p}_1+\vec{p}_2-2\vec{p}_3)/\sqrt{6}$ being the relative coordinates and momenta. $g_d=3/4$ and $g_{t/^3\mathrm{He}}=1/4$ are the statistical factor for spin 1/2 proton and neutron to form a spin 1 deuteron (spin 1/2 triton/$^3$He). The width of the Wigner function is chosen to be $\sigma_d=2.26$ fm, $\sigma_t=1.59$ fm, and $\sigma_{^3\mathrm{He}}=1.76$ fm, consistent with the radius of deuterons, tritons, and $^3$He~\cite{Ropke:2008qk}, respectively. All possible combinations of neutrons and protons are considered in forming light clusters, and this approach has been shown to describe the production of light clusters in both low- ~\cite{Chen:2003ava} and high-energy~\cite{Zhao:2020irc} heavy-ion collisions. The total free spectator neutrons are composed of the residue neutrons that have not coalesced into light clusters and those from the deexcitation of heavy clusters.

\begin{figure}[ht]
\includegraphics[width=0.8\linewidth]{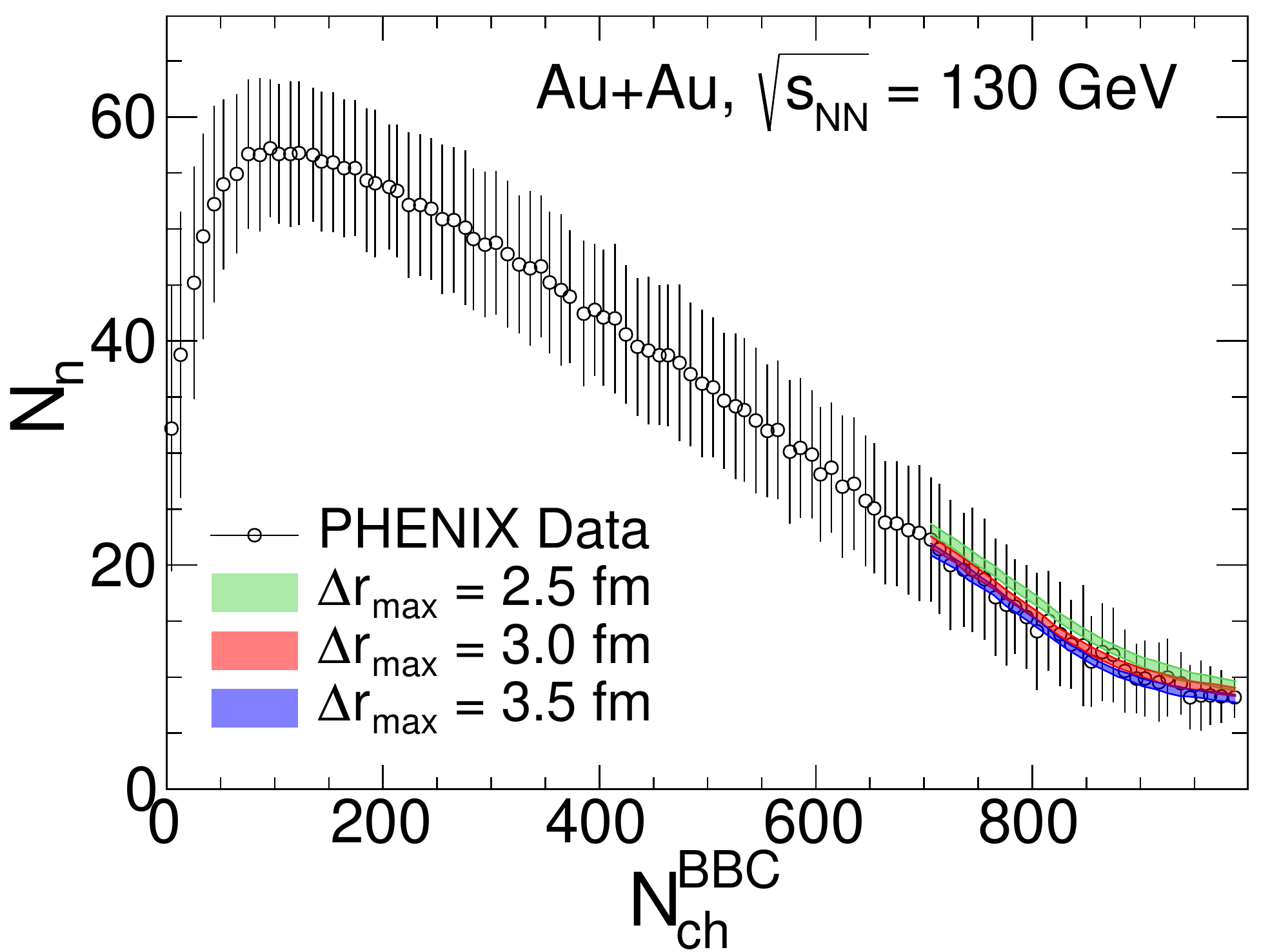}
\caption{Comparison of free spectator neutron numbers for three values of $\Delta r_{\mathrm{max}}$ with the RHIC data~\cite{PHENIX} in Au+Au collisions at $\sqrt{s_{NN}}=130$ GeV, at large charged-particle multiplicities $N_{\mathrm{ch}}^{\mathrm{BBC}}$ measured by the BBC detector at $3.0<|\eta|<3.9$. The error bars on the data represent the RMS width of the correlation. The band for each $\Delta r_{\mathrm{max}}$ indicates the range covered by $L=30$ to 120 MeV.}\label{fig2}
\end{figure}

{\bf Validation with experimental data.} To validate the method and the values of $\Delta r_{\mathrm{max}}$ and $\Delta p_{\mathrm{max}}$ described above, we performed an analysis of Au+Au collisions at $\sqrtsnn=130$ GeV, for which the experimental data on $N_n$ as a function of $\nch$ exists~\cite{PHENIX}. The quadrupole deformation is chosen to be $\beta_{\mathrm{2Au}}=-0.15$~\cite{Giacalone:2021udy,Moller:2015fba} and the nucleon-nucleon inelastic cross section is taken to be 40 mb. The values of $N_n$ as a function of $\nch$ from our calculation are compared with experimental data in Fig.~\ref{fig2}. Here our main focus is on UCC region ($\nch>700$), where the results are not affected by uncertainties of cluster deexcitations and other physics processes.  A smaller $\Delta r_{\mathrm{max}}$ would imply that nucleons in the nuclear surface, enriched with neutrons, can only combine with close-by nucleons to form clusters. Therefore one expect more residue free neutrons, compared to the case with a larger $\Delta r_{\mathrm{max}}$. We see that the value of $\Delta r_{\mathrm{max}}=3$~fm achieves the best description of the experimental data~\cite{PHENIX}. We checked that the results are not sensitive to the value of $\Delta p_{\mathrm{max}}$ within its reasonable range.

\begin{figure}[ht]
\includegraphics[width=0.95\linewidth]{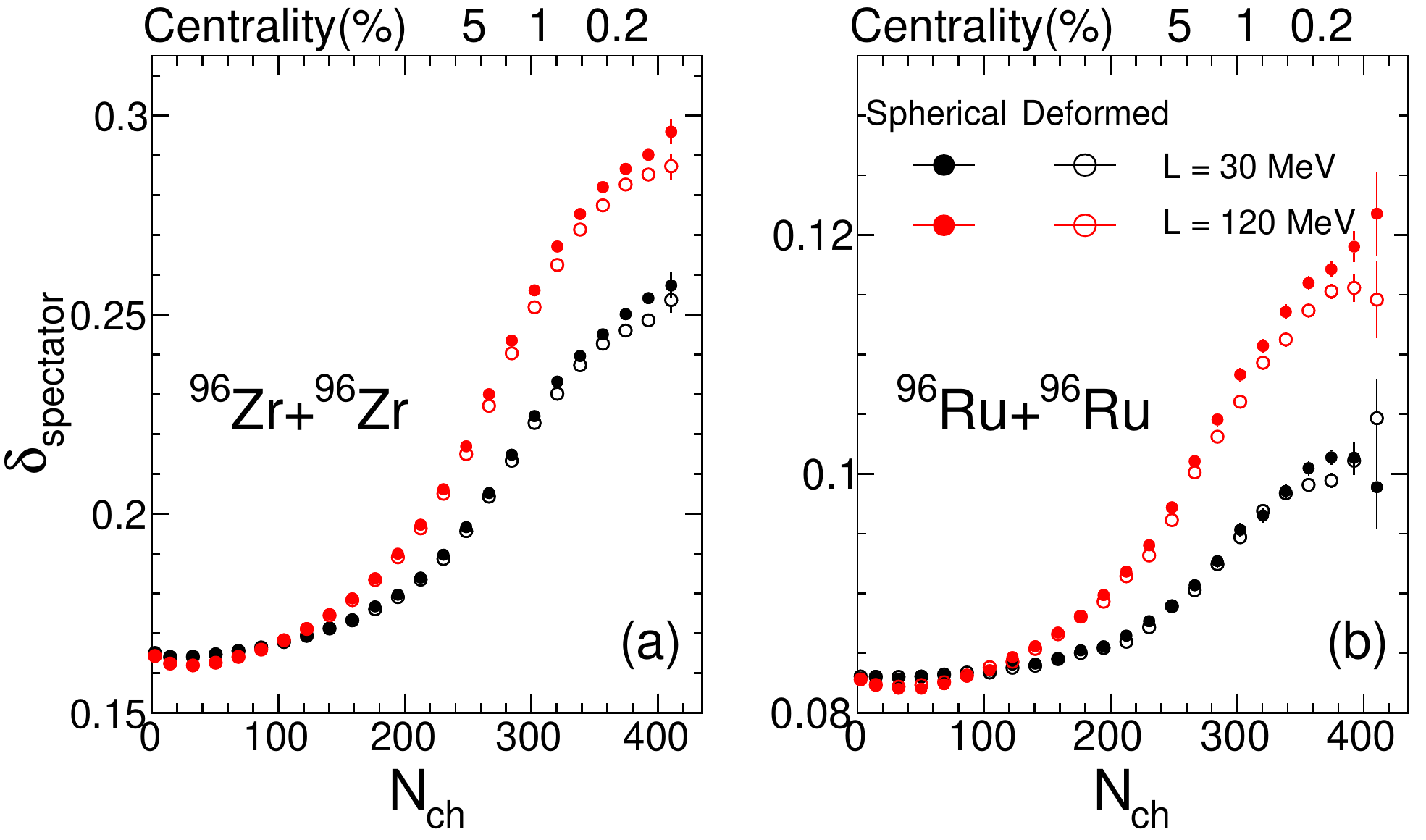}
\caption{Isospin asymmetries of spectator matter in $^{96}$Zr+$^{96}$Zr (a) and $^{96}$Ru+$^{96}$Ru (b) collision systems with density distributions from spherical or deformed SHFB calculations using a smaller or larger $L$.}\label{fig3}
\end{figure}

{\bf Predictions for the isobar systems.} Figure~\ref{fig3} displays the overall isospin asymmetry $\delta_{\mathrm{spectator}}$ of the spectator matter as a function of $\nch$. Due to the presence of neutron skin (see Fig.~\ref{fig1}), the spectator matter becomes more neutron rich in more central collisions. Most of the increase, by about 50\%, happens over the 0--5\% centrality range ($\nch=250-400$), implying that the $\delta_{\mathrm{spectator}}$ in the UCC region is much more sensitive to the neutron skin. Furthermore, the more neutron-rich $^{96}$Zr+$^{96}$Zr system has an overall larger $\delta_{\mathrm{spectator}}$ value than the $^{96}$Ru+$^{96}$Ru system. Most importantly, for a larger $L$ value and therefore a larger neutron skin, the $\delta_{\mathrm{spectator}}$ in central collisions is greatly enhanced. When $L$ increases from 30 to 120 MeV, $\delta_{\mathrm{spectator}}$ in the UCC region increases by about 0.040 in the more neutron-rich $^{96}$Zr+$^{96}$Zr system compared to about 0.015 in the less neutron-rich $^{96}$Ru+$^{96}$Ru system. Inclusion of nuclear deformation reduces slightly the $\delta_{\mathrm{spectator}}$, since the nuclear deformation, combined with random collision orientation, smears the radial distribution of the spectator protons and neutrons. But such effect is subdominant compared to the influence of $L$.

\begin{figure}[ht]
\includegraphics[width=1\linewidth]{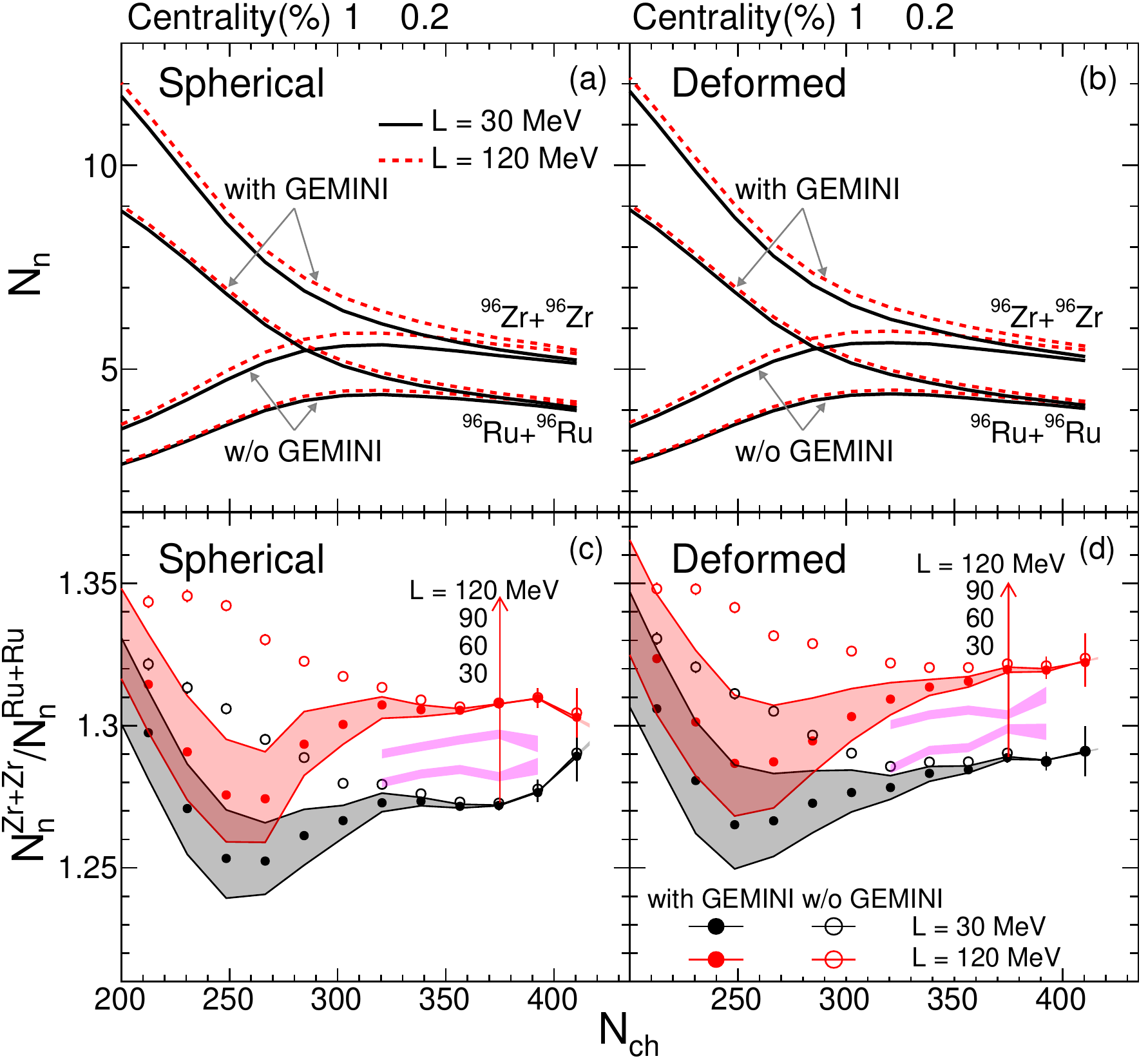}
\caption{Number of free spectator neutrons in $^{96}$Zr+$^{96}$Zr and $^{96}$Ru+$^{96}$Ru collision systems (upper) as well as their ratios (lower) with density distributions from spherical (left) or deformed (right) SHFB calculations using different $L$ values.}\label{fig4}
\end{figure}

Figure~\ref{fig4}\;(a) shows the number of free neutrons $N_n$ with and without considering the deexcitations of heavy clusters in spectator matter via the GEMINI model in the isobar systems. The cluster deexcitations contribute significantly to $N_n$ in non-central collisions but are less important in the UCC region of $\nch>340$ or centrality range $0-0.2\%$, where $N_n$ with and without deexcitations agree with each other, showing that the deexcitation of heavy clusters is no longer the main contribution to $N_n$ there. On the other hand, we note that among spectator nucleons that do not form heavy clusters, about $0.8-1.4$ neutrons, dependent on $\nch$, may coalesce into light clusters. The larger $\Delta r_{\mathrm{np}}$ associated with $L=120$ MeV leads to systematically a larger $N_n$, compared with $L=30$ MeV. The $N_n$ in a single collision system is susceptible to theoretical and experimental uncertainties, such as clusterization algorithm, excitation energy, or detector efficiency, etc. In contrast, the ratio of $N_n$ in ultracentral $^{96}$Zr+$^{96}$Zr to $^{96}$Ru+$^{96}$Ru collisions, as shown in the lower panels of Fig.~\ref{fig4}, is rather insensitive to the uncertainties mentioned above. We predicts a ratio of $N_n$ in the UCC region at around 1.27 for $L =30$~MeV and 1.31 for $L=120$~MeV, respectively. The $L$ effect is not small if the values are compared with the baseline ratio ($ \sim 1.08$) of the total neutron numbers in the two collision systems. The shaded bands in the ratio represent uncertainties from the calculation of the excitation energy ($\pm 1$ MeV per nucleon) of clusters: a higher excitation energy generally leads to more free nucleons. The ratio in the UCC region of $\nch>340$ is clearly insensitive to this uncertainty. Ratios of $N_n$ in the UCC region for $L=60$ and 90 MeV are also plotted in Fig.~\ref{fig4} (c) and (d), which show a linearly increasing trend of the ratio with increasing $L$.

The recent STAR paper~\cite{STAR:2021mii} indeed acknowledged a significant enhancement of $N_n$ in  $^{96}$Zr+$^{96}$Zr collisions relative to $^{96}$Ru+$^{96}$Ru collisions, as measured by the ZDC. It would be exciting to compare our predicted ratio to the future measurement to provide a quantitative constraint on the neutron skin and $L$ value.

The right panels of Fig.~\ref{fig4} show corresponding results including effects of nuclear deformation. In the UCC region of $\nch>340$, the values of $N_n$ are found to decrease by up to $20\%$ compared to the spherical case, and the ratio of $N_n$ changes by about 0.01 when nuclear deformation is enabled, which is about 25\% effect compared to the difference between $L=30$ and 120 MeV. This shows that the potential change to the $N_n$-ratio associated with uncertainties in the deformation is a subleading effect, and is controllable once the deformation is known in prior.

{\bf Summary and outlook.} We show that the yield ratio of free spectator neutrons produced in ultracentral high-energy isobaric collisions, avoiding the complex dynamics for observables at midrapidities, is a clean probe of the neutron-skin thickness $\Delta r_{\mathrm{np}}$ of colliding nuclei and the slope parameter $L$ of the symmetry energy. We illustrate this idea by using a Glauber model for $^{96}$Zr+$^{96}$Zr and $^{96}$Ru+$^{96}$Ru collisions, where proton and neutron distributions in $^{96}$Zr and $^{96}$Ru are provided by the Skyrme-Hartree-Fock-Bogolyubov calculation, and free spectator neutrons are produced either from the deexcitation of heavy clusters in spectator matter or direct ones that have not coalesced into light clusters. The coalescence parameters for the formation of heavy clusters are tuned to describe the measured free spectator neutron yield $N_n$ at large charged-particle multiplicities in Au+Au collisions at $\sqrtsnn=130$ GeV. The values of $N_n$ are predicted to be larger in $^{96}$Zr+$^{96}$Zr than in $^{96}$Ru+$^{96}$Ru collisions due to the larger $\Delta r_{\mathrm{np}}$ in $^{96}$Zr than in $^{96}$Ru. This difference further increases when a larger $L$ value is used. We found that the ratio of $N_n$ in ultracentral $^{96}$Zr+$^{96}$Zr to $^{96}$Ru+$^{96}$Ru collisions are free from the uncertainties of cluster deexcitations and is relatively insensitive to the nuclear deformation. This ratio is also expected to largely cancel experimental errors, therefore making it an clean probe of the $\Delta r_{\mathrm{np}}$ of colliding nuclei.

The present study can be generalized to any two colliding systems with similar mass number but different isospin asymmetries such as those along an isotopic chain, for which the sensitivity of the $N_n$-ratio to $\Delta r_{\mathrm{np}}$ and $L$ is expected to increase with increasing difference in their isospin asymmetries. For a single colliding system, such as Au+Au or Pb+Pb, we found that the yield ratio of free spectator neutrons over free spectator protons is also an excellent probe of $\Delta r_{\mathrm{np}}$ and $L$. The free spectator protons can be measured by instrumenting the forward region with dedicated detectors, see, e.g., Ref.~\cite{Tarafdar:2014oua}. Such study is in progress.

JX and JZ are supported by the National Natural Science Foundation of China under Grant No. 11922514. JJ and CZ are supported by the US Department of Energy under Contract No. DEFG0287ER40331. GXP and LL are supported by the National Natural Science Foundation of China under Grant Nos. 11875052, 11575190, and 11135011.

\bibliography{isobar}
\end{document}